# First-principles exploration of the pressure-dependent physical properties of Sn$_4$Au: a superconducting topological semimetal

M. Abdul Hadi Shah[1,2], M.I. Naher[1], S.H. Naqib[1,*]

[1]Department of Physics, University of Rajshahi, Rajshahi 6205, Bangladesh

[2]Department of Physics, Rajshahi University of Engineering and Technology, Rajshahi 6204, Bangladesh

*Corresponding author: Email: salehnaqib@yahoo.com

**Abstract**

First-principles investigation within the density functional theory is utilized to explore the physical properties of a superconducting topological semimetal Sn$_4$Au under pressure within the range of 0-5 GPa. The structural stability and mechanical stability are justified over the whole pressure range. According to the computed elastic moduli, the compound under study is classified as ductile and applied pressure enhances the ductility, and therefore, escalates its plasticity. The compound has very high level of dry lubricity and machinability index. All the anisotropy factors demonstrate an elastically anisotropic nature, and the strength of anisotropy changes in an anomalous fashion in different contexts, whether pressure is applied or not. The electronic properties are investigated in view of the electronic band structure and density of states. The band structure reveals the topological semimetallic feature of Sn$_4$Au while the density of states at the Fermi level decreases gradually with increasing pressure. Both ionic and covalent bondings are observed in Sn$_4$Au based on the results of Mulliken atomic population analysis and charge density distribution mapping. Optical parameters of Sn$_4$Au are investigated at different pressures. The characteristic peaks in reflectivity, refractive index and photoconductivity exhibit a shift towards higher energy with increasing pressure for all polarizations of the electric field vector. The absorption coefficient and reflectivity spectra designate Sn$_4$Au as a suitable system for optoelectronic applications. For instance, the investigated material might be integrated into infrared detectors due to fairly its high refractive index in the IR region, and it can be used to reduce solar heating in the visible region. Moreover, the pressure dependent shifts in the electronic density of states at the Fermi level, the changes in the Debye temperature, and pressure induced variations in the repulsive Coulomb pseudopotential have been used to explore the effect of pressure on the superconducting transition temperature in this study.

## 1. Introduction

Topological electronic materials, a transdisciplinary and prominent field in condensed matter physics, materials science or even in solid state chemistry [1]. These compounds exhibit unconventional linear response and the evolution of topological states is highly influenced by the topologically invariant properties such as time reversal symmetry or spatial inversion symmetry, reflection symmetry etc. [2]. After the development of electronic topological insulators in 2010s [3–5], theorists have turned their focus on topological semimetals which is a new frontier in the field of contemporary condensed matter physics for not only the fundamental physical behaviors but also for their potential technological implications [6–8].

Topological semimetals (TSMs) can be distinguished from the conventional solids and are characterized by a topologically nontrivial gapless electronic structure originating from topologically stable Fermi surfaces. On the basis of the band crossing/degeneracy, co-dimension (either line or point degeneracy), origin of the crossing and the dispersion in the vicinity of the crossing, TSMs can be categorized in a number of growing families; namely Weyl semimetals (WSMs) [9–12], Dirac semimetals (DSMs) [13–15], nodal line semimetals (NLSMs) [16,17], triple-point semimetals (TPSMs) [18,19], multifold fermion semimetals [20], etc. Fruitful research relating to widespread DSMs and WSMs on a large scale has been conducted during the past decade where band crossings appear along the line rather than the surface in both semimetals. The only two essential criteria to segregate them are crystal structure and magnetism [1]. The appearances of WSMs are possible only if the centrosymmetric crystal systems with magnetism (commonly ferromagnetism) breaks the time-reversal symmetry [12]; whereas Dirac points in DSMs exist in centrosymmetric crystals without magnetism. In WSMs, spin-up and spin-down bands are fully separated due to the presence of magnetism except at high-symmetry points of the Brillouin zone, and Weyl points in WSMs appear with two-fold degeneracy and come in pairs with opposite chirality [21]. In contrast, all the bands exhibit two-fold degeneracy in DSMs according to Kramers' degeneracy [21]. Absence of magnetism results in two-fold bands at the Dirac point, and band crossing introduces four-fold degenerate states [6,8].

Though TSMs do not have a topological order, they still possess topologically protected states that contribute to the development of many exotic properties, making them promising for low-energy electronic device applications [22]. In contrast, superconducting topological semimetals (STSMs), although rare, have become one of the prominent classes in physical sciences [23–25], in which a bulk state demonstrates a superconducting gap while the gapless edge states are composed of Majorana fermions.

Superconductivity with TSMs have been studied recently by some research groups with large magnetoresistance in normal state to realize the surface states [25–28]. Among these, $Sn_4Au$ was found with isomorphic symmetry with those of recently investigated $Sn_4Pd$ [29] and $Sn_4Pt$ TSMs [30]; although the superconducting state of $Sn_4Au$ was investigated in 1960s [31,32].

In recent times, first-principles investigations have been proven to be an instrumental tool in exploring candidate materials for TSMs and played a pivotal role in bridging the theory and experiments by predicting novel topological systems with success [8,33,34].

Recently, M.M. Sharma *et al.* [35] have synthesized Sn$_4$Au having orthorhombic symmetry and found the presence of topological surface states within the system with superconducting properties below a critical temperature ($T_c$) of 2.6 K. Karn *et al.* [36], in the meantime, utilized generalized gradient approximation (GGA) to reveal topological features of AuSn$_4$ with spin–orbit coupling (SOC) and without SOC via electronic band structure calculations. They predicted Sn$_4$Au as a Dirac-type topological semimetal with non-vanishing density of states (DOS) at Fermi level ($E_F$) and found the negligible but not completely redundant effect of the SOC. Earlier, in 1984 and 1985, Rubiak *et al.* [37,38] experimentally measured the unit cell and positional parameters within the structure refinement procedure of Sn$_4$Au and PdSn$_4$.

Apart from the structural and electronic features at ambient pressure, no further studies are found in the scientific literature on physical properties of Sn$_4$Au. Moreover, theoretical investigations of the system under hydrostatic pressure do not exist; further studies are therefore necessary on a large scale to explore the physical properties of Sn$_4$Au semimetal. As known, experimental investigation of the pressure-dependent physical properties is a challenging task. On the contrary, the theoretical approach, such as the first-principles method based on density functional theory (DFT), would be a quite reliable tool in understanding physical properties under pressure [39,40]. Pressure is also a clean tool to understand the change in the ground state physical properties as the crystal volume is changed. This is particularly important for systems with unconventional electronic band structure, topological order, and superconductivity [41-45].

Therefore, in the work presented herein, we intend to probe the pressure-induced physical behavior of Sn$_4$Au semimetal by means of structural evolution, elasto-mechanical features along with anisotropic nature in view of elastic moduli, elasto-acoustic properties, hardness, electronic band structure, and optical properties using the DFT framework. As the coexistence of superconductivity and nontrivial band topology is rare; materials with dual nature would provide enlightenment for further theoretical and experimental investigations to find comparable prototype novel materials.

This manuscript is organized further in three main parts. The methodology of calculations together with relevant equations is mentioned in 'Computational scheme' section. 'Results and discussion' section comprises the findings regarding various physical properties of Sn$_4$Au, and finally all of the key findings are summarized in the 'Conclusion' section.

## 2. Computational scheme

In this study, the first-principles calculations have been carried out based on DFT [46] as implemented within the CASTEP [47]. The solids-corrected Perdew-Burke-Ernzerhof (PBEsol) functional [48] in the generalized gradient approximation (GGA) is adopted for exchange-correlations (XC), while the Vanderbilt-type ultrasoft pseudopotential is employed for the interaction of charges between atomic core and valence electrons. The further corrected GGA method by Perdew *et al.* [49] is capable of yielding more precise physical properties for densely packed solids. To ensure superior convergence of the total energy, a plane-wave basis set cut-off of 550 eV and 108 irreducible *k*-points through Monkhorst-Pack grid of 11×11×6 was selected in the Brillouin zone integrations. The equilibrium structural optimization was carried out using the Broyden–Fletcher–Goldfarb–Shanno (BFGS) minimization algorithm [50] for both ground state and hydrostatic pressures conditions. Pulay density mixing scheme was adopted for the electron energy minimization process. A self-consistent field tolerance of $1.0 \times 10^{-6}$ eV/atom was set for total energy estimation. The total energy of the system, interaction force between the atoms, stress and the maximum displacement converges to $1.0 \times 10^{-5}$ eV/atom, 0.03 eV/Å, 0.05 GPa and $1.0 \times 10^{-3}$ Å, respectively, during optimization process.

The single-crystal elastic constants for $Sn_4Au$ are estimated within the 'stress–strain' method [51] embedded in the CASTEP package. In the case of the orthorhombic structure, there are nine independent elastic stiffness constants $C_{ij}$, *viz*, $C_{11}$, $C_{12}$, $C_{13}$, $C_{22}$, $C_{23}$, $C_{33}$, $C_{44}$, $C_{55}$ and $C_{66}$. For a stable orthorhombic structure, all nine elastic constants should satisfy the following necessary and sufficient conditions [52,53]:

$$\left.\begin{array}{r} C_{11} > 0;\ C_{11}C_{22} > C_{12}^2; \\ C_{11}C_{22}C_{33} + 2C_{12}C_{13}C_{23} - C_{11}C_{23}^2 - C_{22}C_{13}^2 - C_{33}C_{12}^2) > 0; \\ C_{44} > 0;\ C_{55} > 0;\ C_{66} > 0 \end{array}\right\} \quad (1)$$

Furthermore, the mechanical stability criterion under hydrostatic compression for the same can be written as [54]:

$$\left.\begin{array}{r} \tilde{c}_{11} + \tilde{c}_{22} - 2\tilde{c}_{12} > 0;\ \tilde{c}_{11} + \tilde{c}_{33} - 2\tilde{c}_{13} > 0;\ \tilde{c}_{22} + \tilde{c}_{33} - 2\tilde{c}_{23} > 0; \\ \tilde{c}_{ii} > 0\ (i = 1 \sim 6); \\ \tilde{c}_{11} + \tilde{c}_{22} + \tilde{c}_{33} + 2\tilde{c}_{12} + 2\tilde{c}_{13} + 2\tilde{c}_{23} > 0. \\ where, \tilde{c}_{ii} = c_{ii} - P\ (i = 1 \sim 6);\ \tilde{c}_{12} = c_{12} + P;\ \tilde{c}_{13} = c_{13} + P;\ \tilde{c}_{23} = c_{23} + P \end{array}\right\} \quad (2)$$

The polycrystalline bulk and shear moduli (*B* and *G*) can be estimated by using $C_{ij}$ by means of the Voigt-Reuss-Hill scheme [55,56]. In this scheme, *B* and *G* are estimated by arithmetic average of Voigt and Reuss bounds as:

$$G = \frac{G_V + G_R}{2};\ B = \frac{B_V + B_R}{2} \quad (3)$$

The Voigt shear and bulk modulus ($G_V$ and $B_V$) for orthorhombic lattices are calculated directly from the elastic constants, $C_{ij}$ as [57-59]:

$$G_V = \frac{1}{15}[(C_{11} + C_{22} + C_{33}) - (C_{12} + C_{13} + C_{23}) + 3(C_{44} + C_{55} + C_{66})]$$
$$B_V = \frac{1}{9}[(C_{11} + C_{22} + C_{33}) + 2(C_{12} + C_{13} + C_{23})]$$
(4)

Subsequently, various elasto-mechanical parameters viz. Young's modulus (*Y*), Poisson's ratio (*σ*), Kleinman parameter (*ξ*), and Lamé's coefficients (*λ* and *μ*) are estimated as follows [60]:

$$Y = \frac{9BG}{3B + G} \text{ and } \sigma = \frac{3B - 2G}{2(3B + G)} \quad (5)$$

$$\xi = \frac{C_{11} + 8C_{12}}{7C_{11} + 2C_{12}} \quad (6)$$

$$\lambda = \frac{Y\sigma}{(1 + \nu)(1 - 2\nu)} \text{ and } \mu = \frac{Y}{2(1 + \nu)} \quad (7)$$

The optical features of materials are characterized by their frequency/energy dependent dielectric function *ε(ω)*. This function is a complex tensor for anisotropic materials that explains the linear optical response of an electronic system to the electromagnetic radiation. Conventionally, the dielectric function is written as *ε(ω) = ε₁(ω) + iε₂(ω)*, where *ε₁(ω)* and *ε₂(ω)* correspond to the real and imaginary parts of the dielectric constants, respectively.

The frequency-dependent imaginary part of the dielectric function *ε₂(ω)* indicate the absorption of the incident radiations and is expressed by [61]:

$$\varepsilon_2(\omega) = \left(\frac{e^2\hbar}{\pi m^2 \omega^2}\right) \sum_{v,c} \int_{BZ} |M_{cv}(k)|^2 \delta[\omega_{cv}(k) - \omega] d^3k \quad (8)$$

where $M_{cv}(k) = \langle u_{ck}|\delta\nabla|u_{vk}\rangle$ is the momentum dipole matrix components for direct transitions in between valence $u_{vk}(r)$ and conduction band $u_{ck}(r)$ electrons with the wave vector *k*, and $\hbar\omega_{cv}(k) = (E_{ck} - E_{vk})$ corresponds to the transition energy. The integral is taken over the first Brillouin zone (BZ).

Real part of the dielectric constant *ε₁(ω)*, on the contrary, describes the polarization and can be derived from the imaginary part *ε₂(ω)* using the Kramers-Kronig relation [62]:

$$\varepsilon_1(\omega) = 1 + \frac{2}{\pi} P \int_0^\infty \frac{\omega' \varepsilon_2(\omega')}{\omega'^2 - \omega^2} d\omega' \quad (9)$$

where *P* signifies the principal value of the integral.

To acquire deeper understanding of optical nature of a system, the complex refractive index *N(ω)* is to be estimated. Being a complex quantity, it has two parts: the real part, referred to as refractive index $n(\omega)$ and the imaginary part, known as the extinction coefficient $k(\omega)$. $N(\omega)$ [$= n(\omega) + ik(\omega)$] is calculated using the following expressions [63,64]:

$$n(\omega) = \frac{1}{\sqrt{2}}\left[\sqrt{\varepsilon_1(\omega)^2 + \varepsilon_2(\omega)^2} + \varepsilon_1(\omega)\right]^{1/2} \quad (10)$$

$$k(\omega) = \frac{1}{\sqrt{2}}\left[\sqrt{\varepsilon_1(\omega)^2 + \varepsilon_2(\omega)^2} - \varepsilon_1(\omega)\right]^{1/2} \quad (11)$$

The absorption coefficient, $\alpha(\omega)$ and optical conductivity, $\sigma(\omega)$ can be expressed in terms of $\varepsilon_1(\omega)$ and $\varepsilon_2(\omega)$ as [65]:

$$\alpha(\omega) = \sqrt{2}\omega\left[\sqrt{\varepsilon_1(\omega)^2 + \varepsilon_2(\omega)^2} - \varepsilon_1(\omega)\right]^{1/2} \quad (12)$$

$$\sigma(\omega) = \frac{\omega}{4\pi}\varepsilon_2(\omega) \quad (13)$$

The reflectivity, $R(\omega)$ predicts how much light is reflected from a surface, which can be estimated from the equation [65]:

$$R(\omega) = \left|\frac{\sqrt{\varepsilon(\omega)} - 1}{\sqrt{\varepsilon(\omega)} + 1}\right|^2 = \left|\frac{\sqrt{\varepsilon_1(\omega) + i\varepsilon_2(\omega)} - 1}{\sqrt{\varepsilon_1(\omega) + i\varepsilon_2(\omega)} + 1}\right|^2 \quad (14)$$

The energy-loss function, $L(\omega)$ explains the interaction to estimate the amount of energy lost by a fast moving electron traversing through the material and is calculated as [65]:

$$L(\omega) = \frac{\varepsilon_2(\omega)}{\varepsilon_1^2(\omega) + \varepsilon_2^2(\omega)} \quad (15)$$

## 3. Results and analysis

### 3.1. Structure and stability

The arrangement of atoms and hence structural parameters are predominant concerns in predicting the structure and stability of a solid. Also, structural properties play a decisive role in determining the optimized lattice constants which helps to calculate the other physical properties. $Sn_4Au$, isomorphous to $PtSn_4$ and $PdSn_4$ [38,66], belongs to orthorhombic structure having the space group *Aea2* (41) [25,35,37,38] that contains four formula unit in which Sn and Au atoms in elementary cell are located at Sn1 (0.1694, 0.3395, 0.1242), Au (0, 0, 0) and Sn2 (0.3502, 0.1624, 0.8574) sites of Wyckoff coordinates. The crystal structures of $Sn_4Au$ in real space in 2D and 3D view are generated using the VESTA software, as shown in **Fig. 1**.

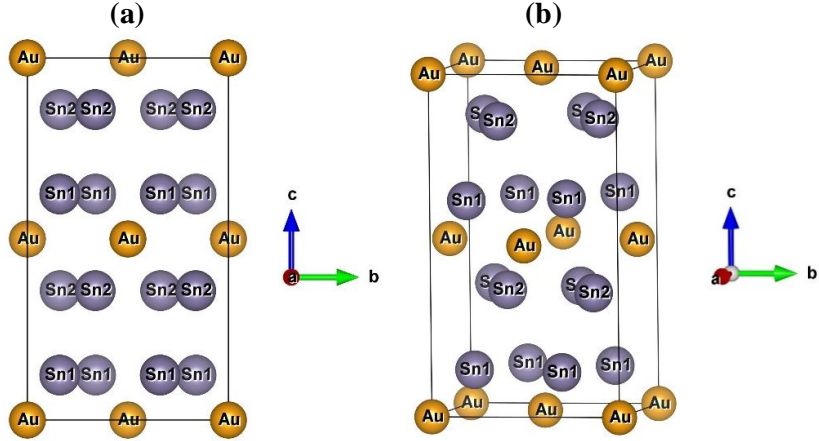

**Fig. 1**. Unit cell structure of Sn$_4$Au in real space: (a) 2D view and (b) 3D view.

To acquire least-deviated equilibrium lattice constants and corresponding cell volume from experimental results, we have considered various exchange-correlation functionals and the estimated results along with the available prior data are summarized in **Table 1**. It is observed that GGA-PBESol provides the minimum deviation of volume from the experimental value among all functionals; thereafter, this functional is used to study various physical properties of Sn$_4$Au in the subsequent sections.

**Table 1**. Calculated equilibrium lattice parameters ($a$, $b$, $c$ in Å), volume ($V$ in Å$^3$), density of crystal ($\rho$ in gm/cm$^3$) and formation energy ($E_f$ in eV/atom) of Sn$_4$Au alongside available experimental data.

| Compound | $a$ | $b$ | $c$ | $V$ | $P$ | $E_f$ | Functionals | Refs. |
|---|---|---|---|---|---|---|---|---|
| Sn$_4$Au | 6.617 | 6.628 | 12.001 | 526.721 | 8.447 | -3.852 | GGA-PBE | This work |
| | 6.639 | 6.653 | 11.993 | 529.702 | 8.423 | -3.524 | GGA-RPBE | This work |
| | 6.611 | 6.616 | 12.161 | 531.914 | 8.388 | -4.001 | GGA-PW91 | This work |
| | 6.348 | 6.358 | 11.513 | 464.667 | 9.602 | -4.780 | LDA | This work |
| | 6.576 | 6.587 | 11.928 | 516.700 | 8.635 | -4.264 | GGA-PBESol | This work |
| | 6.515 | 6.529 | 11.726 | 498.783 | 8.945 | - | Expt. | [Ref][a] |
| | 6.512 | 6.516 | 11.707 | 496.778 | - | - | Expt. | [Ref][b] |

[a][35], [b][37]

**Figure 2** shows the variation of total energy and volume with pressure for Sn$_4$Au semimetal. It is clearly observed that the total energy of the cell increases with pressure (**Fig. 2a**) and decreases with volume (**Fig. 2b**). These behaviors are consistent with the literatures [67,68]. After structural optimization under different pressures, the elasto-mechanical and optoelectronic behaviors at different pressures have been investigated.

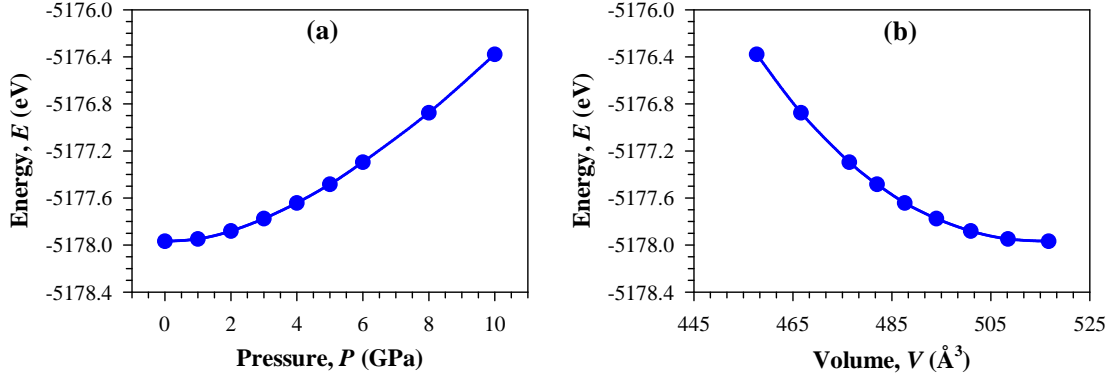

**Fig. 2.** (a) Pressure and (b) volume-dependent total energy for $Sn_4Au$ semimetal.

Pressure-dependent normalized lattice parameters ($a/a_0$, $b/b_0$, $c/c_0$), densities ($\rho/\rho_0$) and volume of the cell ($V/V_0$) for $Sn_4Au$ are illustrated in **Fig. 3**. The calculated lattice constants (**Fig. 3a**) and their volume gradually decrease as pressure rises; meanwhile the densities increase (**Fig. 3b**). This expected findings are again compatible with [67,68]. It is clearly seen that the rate of decrement of $c/c_0$ is higher than $a/a_0$ and $b/b_0$; $a/a_0$ and $b/b_0$ maintain almost the same rate. This behavior signifies that $c$-axis is easily compressible in $Sn_4Au$.

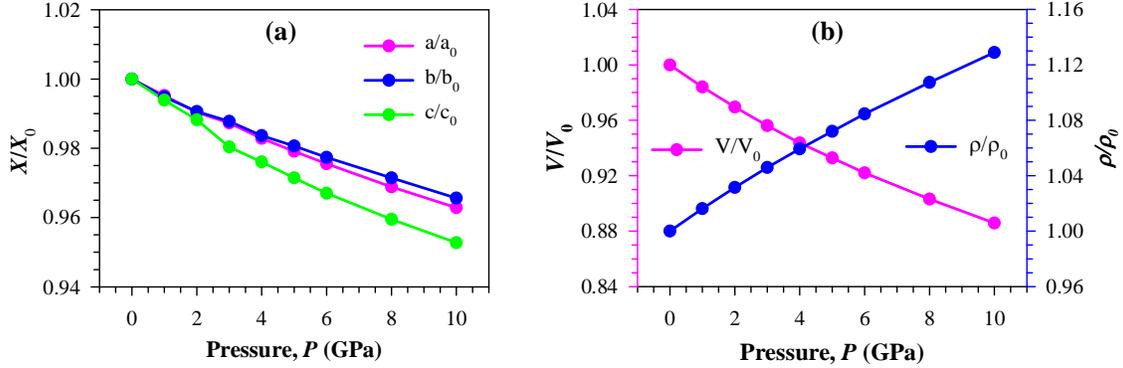

**Fig. 3.** Variation of normalized (a) lattice parameter ($a/a_0$, $b/b_0$, $c/c_0$), and (b) densities ($\rho/\rho_0$) and volume ($V/V_0$) with hydrostatic pressure for the $Sn_4Au$ semimetal. The index 'zero' designates the value in the ground state.

To ensure the thermodynamic stability of $Sn_4Au$, its formation energy ($E_f$) needs to be calculated. In general, negative formation energy implies a stable chemical phase [69]. The formation energy is estimated as:

$$E_f = \frac{1}{a+b}\left[E_{Sn_4Au}^{total} - \left(aE_{Sn}^{bulk} + bE_{Au}^{bulk}\right)\right] \quad (16)$$

where $E_{Sn_4Au}^{total}$ represents the total energy of $Sn_4Au$ per formula unit, and $E_{Sn}^{bulk}$ and $E_{Au}^{bulk}$ refer to the ground state energies of Sn and Au in the bulk state, respectively; $a$ and $b$ represents the corresponding number of Sn and Au atoms in the cell. The calculated values of $E_f$ for various functionals and their pressure-dependent variations are presented in **Table 1** and **Fig. 4**, respectively. The results recommend their chemical stability at pressures up to 10 GPa. Moreover, the more negative $E_f$ indicates the better thermodynamic stability of compounds [70]. Therefore, it is evident that the stability weakens as we go from ambient to higher pressures (**Fig. 4**).

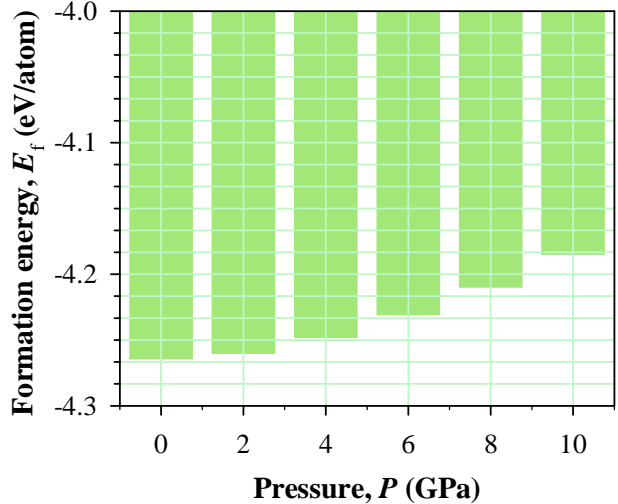

**Fig. 4**. The formation energies/atom, ($E_f$), of Sn$_4$Au as a function of pressure.

*3.2 Elastic constants and mechanical properties*

Elastic constants play a decisive role in providing crucial information on the mechanical and dynamical properties of crystals. The second-order single-crystal elastic coefficients of Sn$_4$Au, as well as related polycrystalline features such as bulk modulus ($B$), shear modulus ($G$), and Young's modulus ($Y$), are computed under ambient and hydrostatic pressure conditions up to 5 GPa with an interval of 1 GPa. The calculated single crystal and polycrystalline elastic parameters at ambient condition are tabulated in **Table 2**; pressure-dependent variations are shown in **Fig. 5**. The mechanical stability of Sn$_4$Au is examined at ambient and under pressure by using the modified Born-Huang stability criteria (**Eqns. 1** and **2**). It is evident that the compound is mechanically stable up to 5 GPa as it satisfies all the stability criteria. We have found a sharp decrease in $C_{33}$ close to 3 GPa, implying that the crystal becomes anomalously more compressible in the *c*-direction around this particular pressure.

Elastic stiffness constants $C_{11}$, $C_{22}$ and $C_{33}$ describes the strength of atomic bonding along the *a*-, *b*-, and *c*-axis of the unit cell, respectively; while, $C_{44}$, $C_{55}$ and $C_{66}$ explains the resistance to the shear deformation involving different crystal planes and directions [71]. The off-diagonal elastic constants, $C_{12}$, $C_{13}$ and $C_{23}$ are also linked with the resistances to shearing strains. In contrast, polycrystalline bulk, shear and Young's moduli of materials reveal the bulk resistance to change in volume, shape and materials stiffness, respectively [72]. It is observed from **Fig. 5a** that $C_{11} \sim C_{22} > C_{33}$, suggesting that atomic bonds along in the basal plane are stronger than those along the *c*-direction. It is noteworthy that the isotropic bulk modulus is always higher than the shear modulus, indicating less pronounced directional bonding within the atoms [71] of $Sn_4Au$. It is to be highlighted that there are no available results either experimental or theoretical regarding the elastic parameters of $Sn_4Au$, and this work can serve as a valuable source of reference for upcoming researches. All the elastic constants tend to increase with pressure at different rates (standard behavior), except $C_{33}$ close to 3 GPa.

**Table 2**. Calculated single crystal and polycrystalline elastic parameters ($C_{ij}$, $B$, $G$, $Y$ all in GPa) for $Sn_4Au$ in the ground state.

| Compound | $C_{11}$ | $C_{12}$ | $C_{13}$ | $C_{22}$ | $C_{23}$ | $C_{33}$ | $C_{44}$ | $C_{55}$ | $C_{66}$ | $B$ | $G$ | $Y$ |
|---|---|---|---|---|---|---|---|---|---|---|---|---|
| $Sn_4Au$ | 92.787 | 42.317 | 47.267 | 92.743 | 51.475 | 79.876 | 12.275 | 8.209 | 33.517 | 60.785 | 17.011 | 46.678 |

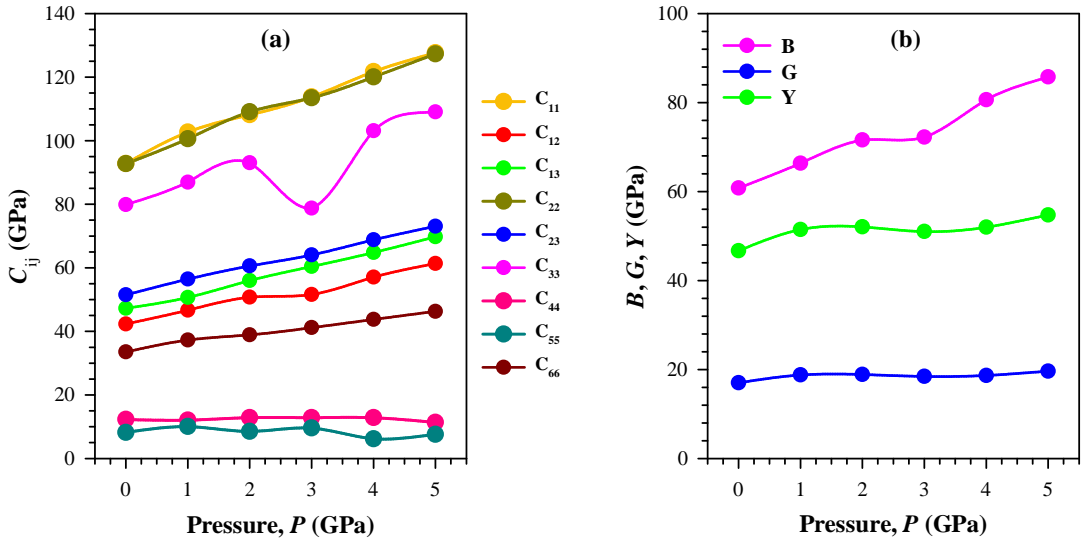

**Fig. 5**. Variation of (a) elastic constants $C_{ij}$, and (b) elastic moduli ($B$, $G$, $Y$) with hydrostatic pressure of $Sn_4Au$.

**Table 3**. Calculated Poisson's ratio ($\sigma$), Pugh's ratio ($G/B$), Kleinman parameter ($\zeta$), machinability index ($\mu_M$), Cauchy pressure [($C_{23}$-$C_{44}$), ($C_{13}$-$C_{55}$) and ($C_{12}$-$C_{66}$) in GPa], bulk modulus ($B_a$, $B_b$, and $B_c$ in GPa) along '*a*', '*b*' and '*c*' axis, isotropic Bulk modulus ($B_i$) and Vickers hardness ($H_V$ in GPa) for $Sn_4Au$ at 0 GPa.

| Compound | $\sigma$ | $G/B$ | $\zeta$ | $\mu_M$ | $C_{23}$-$C_{44}$ | $C_{13}$-$C_{55}$ | $C_{12}$-$C_{66}$ | $B_a$ | $B_b$ | $B_c$ | $B_i$ | $H_V$ (Tian) | $H_V$ (Teter) | $H_V$ (Miao) | $H_V$ (Mazhnik) |
|---|---|---|---|---|---|---|---|---|---|---|---|---|---|---|---|
| $Sn_4Au$ | 0.372 | 0.28 | 0.62 | 4.95 | 39.20 | 39.06 | 8.80 | 185.1 | 204.4 | 162.1 | 60.7 | 1.61 | 2.57 | 1.45 | 2.67 |

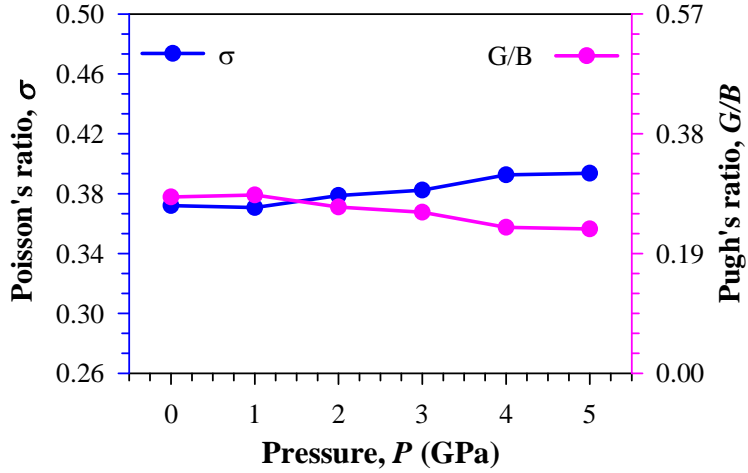

**Fig. 6**. Pressure-dependent Poisson's ratio ($\sigma$) and Pugh's ratio ($G/B$) of Sn$_4$Au semimetal.

Poisson's ratio ($\sigma$) of solids quantifies the change in volume during uniaxial deformation; its maximum value of 0.5 means no change in volume during elastic deformation [57]. It also forecasts the ductility/brittleness, plasticity and nature of bonding forces [73] in the crystal. A material is classified as ductile if $\sigma > 0.26$, otherwise, it is brittle [74]. The bonding forces between the atoms are classified as central if $\sigma$ lies within 0.25 to 0.50 [75]. The evaluated values of $\sigma$ range from 0.362 to 0.394 in the pressure range 0-5 GPa, suggesting that Sn$_4$Au is highly ductile in nature. Moreover, the fairly high values (substantially higher than 0.25) of $\sigma$ categorize it as a better plastic material [76,77] and indicates a small volume change associated with its deformation [57]. Pressure-dependent Poisson's ratio of Sn$_4$Au (as seen in **Fig. 6**) exhibits that applied pressure enhances the ductility and therefore, escalates its plasticity. Also, a predominant central interatomic force between the atoms appears in Sn$_4$Au within the whole range of pressure considered. The high $\sigma$ also indicates that the bonding is less directional and fairly central in nature.

Another parameter utilized for distinguishing ductility and brittleness of solids is the Pugh's ratio ($G/B$) [78]. The critical value of ($G/B$) is approximately 0.57. If ($G/B$) is less than 0.57, the material is ductile, otherwise it is brittle. The variation of ($G/B$) with pressure for Sn$_4$Au up to 5 GPa (shown in **Fig. 6**) indicates ductile behavior throughout the pressure range; ductility increasing as pressure rises.

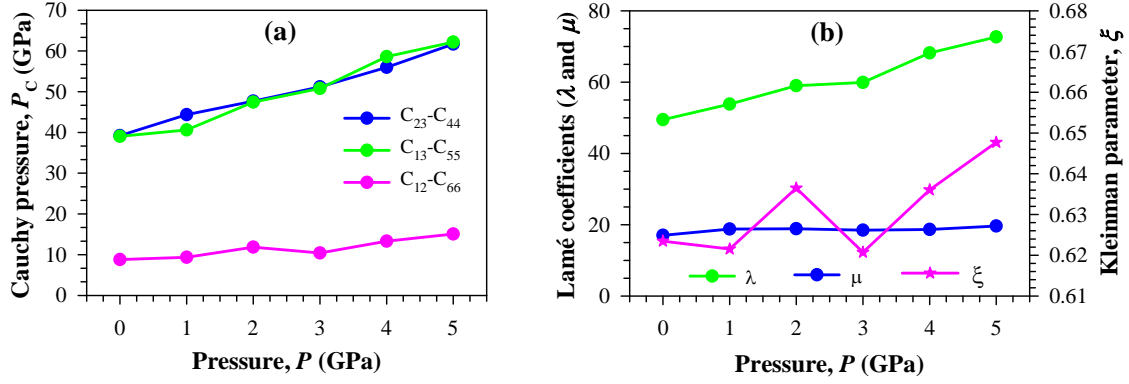

**Fig. 7**. Pressure-dependent (a) Cauchy pressure and (b) Lame coefficients and Kleinman parameter of $Sn_4Au$.

Cauchy pressure ($P_C$) can also judge the bonding nature of solids based on the negative or positive values of $P_C$, according to Pettifor [79]. In orthorhombic crystals, $P_C$'s are defined as ($C_{23}-C_{44}$), ($C_{13}-C_{55}$), and ($C_{12}-C_{66}$). $P_C$ is typically positive for metallic (damage-tolerant) bonding; otherwise it represents directional bonding with angular behavior [80]. The pressure-dependent Cauchy pressures of $Sn_4Au$ are depicted in **Fig. 7a**. It is evident that all Cauchy pressures, with and without pressure, are positive. This implies the dominance of a ductile character with metallic bonding in $Sn_4Au$.

An important mechanical index, known to be the Kleinman parameter $\xi$, judges the internal strain of materials and therefore implies the relative strength of bond bending with respect to bond stretching. The minimum bond stretching and bond bending occurs at $\xi = 1$ and $\xi = 0$, respectively [81]. It is estimated by using **Eqn. 6** and its pressure dependent nature is depicted in **Fig. 7b**. The estimated $\xi$ for $Sn_4Au$ at ambient pressure confer the minimum value of 0.611 (**Table 3**), and therefore, the highest mechanical durability is achieved at 0 GPa pressure because materials are more resistant to changes in bond length compared to higher pressures.

First and second Lamé's constants ($\lambda$ and $\mu$) imply the compressibility and shear stiffness of materials, respectively [82]. These are calculated using **Eqn. 7** and the variation with pressure is plotted in **Fig. 7b**. A material is expected to be isotropic when $\lambda = C_{12}$ and $\mu = (C_{11}-C_{12})/2$ [83]. The present calculation provides a mismatch with these criteria for $Sn_4Au$, suggesting their anisotropic nature with and without applied pressure.

In material's engineering, machinability index ($\mu_M = B/C_{44}$) quantifies material's dry lubricating property and the level of ease of its machining. A high value of $\mu_M$ is desired for efficient manufacturing [84]. Very high value of $\mu_M$ signifies better dry lubricity of $Sn_4Au$ and the application of pressure enhances the effect further as depicted in **Fig. 8a**.

The directional bulk moduli along $a$-, $b$- and $c$-axis and isotropic bulk modulus ($B_i$) are estimated by employing the expressions [85]:

$$B_a = a\frac{dP}{da} = \frac{\Lambda}{1 + \alpha + \beta}; \quad B_b = b\frac{dP}{db} = \frac{B_a}{\alpha}; \quad B_c = c\frac{dP}{dc} = \frac{B_a}{\beta} \tag{17}$$

$$B_i = \frac{\Lambda}{(1 + \alpha + \beta)^2} \tag{18}$$

with,

$$\left.\begin{aligned}\Lambda &= C_{11} + 2C_{12}\alpha + C_{22}\alpha^2 + 2C_{13}\beta + C_{33}\beta^2 + 2C_{23}\alpha\beta \\ \alpha &= \frac{(C_{11} - C_{12})(C_{33} - C_{13}) - (C_{23} - C_{13})(C_{11} - C_{13})}{(C_{33} - C_{13})(C_{22} - C_{12}) - (C_{13} - C_{23})(C_{12} - C_{23})} \\ \beta &= \frac{(C_{22} - C_{12})(C_{11} - C_{13}) - (C_{11} - C_{12})(C_{23} - C_{12})}{(C_{22} - C_{12})(C_{33} - C_{13}) - (C_{12} - C_{23})(C_{13} - C_{23})}\end{aligned}\right\} \tag{19}$$

**Fig. 8a** shows the pressure-induced directional bulk moduli $B_a$, $B_b$ and $B_c$, demonstrating an increasing trend with rising pressure, except close to 3 GPa, where the behavior becomes nonmonotonic.

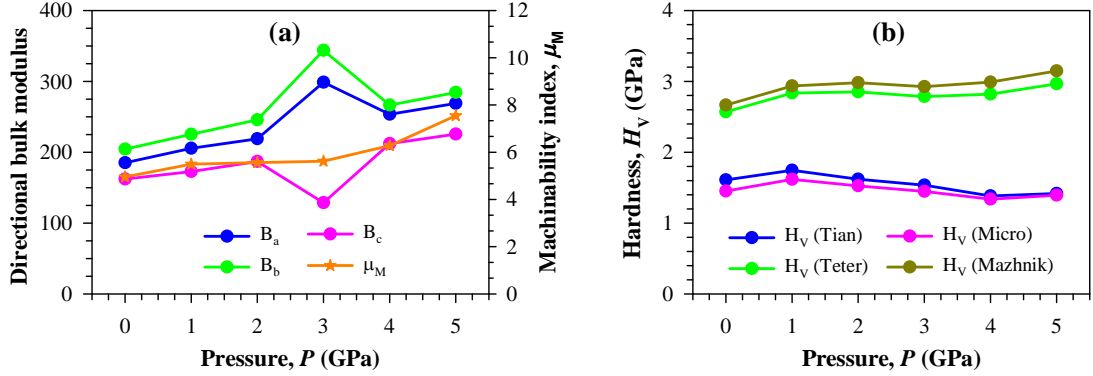

**Fig. 8**. (a) Directional bulk modulus and machinability index $\mu_M$ and (b) Vickers harnesses as a function of pressure for $Sn_4Au$ semimetal.

The theoretical hardness, usually called Vickers hardness ($H_V$), is another important mechanical performance parameter. Among various theoretical schemes for calculating Vickers hardness, four most reliable methods are due to Y. Tian *et al*. [86], D.M. Teter [87], N. Miao *et al*. [88] and E. Mazhnik [89]. Different formalisms are given below:

$$\left.\begin{aligned}(H_V)_{Teter} &= 0.151G \\ (H_V)_{Tian} &= 0.92\left(\frac{G}{B}\right)^{1.137}G^{0.708} \\ (H_V)_{Miao} &= \frac{(1 - 2\sigma)Y}{6(1 + \sigma)} \\ (H_V)_{Mazhnik} &= \gamma_0\chi(\sigma)Y \\ \text{where,}\; \chi(\sigma) &= \frac{1 - 8.5\sigma + 19.5\sigma^2}{1 - 7.5\sigma + 12.2\sigma^2 + 19.6\sigma^3}\end{aligned}\right\} \tag{20}$$

$\gamma_0$ is a dimensionless constant having a value of 0.096.

In the ground state, the estimated values of hardness of $Sn_4Au$ are presented in **Table 3**, and their pressure-dependent variation is plotted in **Fig. 8b**. The estimated values of $H_V$ at 0 GPa are 1.61 GPa, 2.57 GPa, 1.45 GPa and 2.67 GPa for Tian, Teter, Miao and Mazhnik approaches, respectively. It follows the following sequence at each and every studied pressure: $H_V$ (Miao) < $H_V$ (Tian) < $H_V$ (Teter) < $H_V$ (Mazhnik). The average hardness of $Sn_4Au$ is around 2.0 GPa. This classifies the compound as fairly soft.

*3.3 Anisotropy in elasticity*

Estimation of anisotropy in elasticity is crucial in materials science due to its correlation with mechanical strength, micro-crack formation, and plastic deformation development [90]. To get deeper insight of $Sn_4Au$ as a durable and potentially usable material under hydrostatic pressures, this section is aimed to estimate various anisotropy indices *viz.*, shear anisotropy factors ($A_1$, $A_2$, $A_3$), universal anisotropy factor ($A^U$), and percentage anisotropies in compressibility ($A_B$) and shear ($A_G$) etc.

We compute the anisotropies in the bulk modulus for orthorhombic $Sn_4Au$ along *a*- and *c*-axes in relation to *b*-axis ($A_{Ba}$ and $A_{Bc}$) using the following equations [91,92]:

$$A_1 = \frac{4C_{44}}{C_{11} + C_{33} - 2C_{13}}; \quad A_2 = \frac{4C_{55}}{C_{22} + C_{33} - 2C_{23}}; \quad A_3 = \frac{4C_{66}}{C_{11} + C_{22} - 2C_{12}} \quad (21)$$

$$A_B = \frac{B_V - B_R}{B_V + B_R}; \quad A_G = \frac{G_V - G_R}{G_V + G_R} \quad (22)$$

$$A^U = \frac{B_V}{B_R} + 5\frac{G_V}{G_R} - 6 \quad (23)$$

$$A_{Ba} = \frac{B_a}{B_b} = \alpha; \quad A_{Bc} = \frac{B_c}{B_b} = \frac{\alpha}{\beta} \quad (24)$$

**Table 4**. Computed shear anisotropic factors ($A_1$, $A_2$ and $A_3$), percentage anisotropy factors ($A_B$ and $A_G$ in %), universal anisotropy factor $A^U$ and compressibility anisotropy factors ($A_{Ba}$ and $A_{Bc}$) for $Sn_4Au$ at various pressures.

| P (GPa) | $A_1$ | $A_2$ | $A_3$ | $A_B$ (%) | $A_G$ (%) | $A^U$ | $A_{Ba}$ | $A_{Bc}$ |
|---|---|---|---|---|---|---|---|---|
| 0 | 0.628 | 0.471 | 1.329 | 0.084 | 12.224 | 1.394 | 0.906 | 0.793 |
| 1 | 0.547 | 0.538 | 1.355 | 0.112 | 11.771 | 1.336 | 0.912 | 0.767 |
| 2 | 0.577 | 0.420 | 1.344 | 0.116 | 14.361 | 1.679 | 0.891 | 0.760 |
| 3 | 0.717 | 0.598 | 1.327 | 1.244 | 15.921 | 1.919 | 0.869 | 0.375 |
| 4 | 0.539 | 0.290 | 1.371 | 0.081 | 22.360 | 2.882 | 0.951 | 0.796 |
| 5 | 0.468 | 0.337 | 1.400 | 0.081 | 20.736 | 2.618 | 0.946 | 0.793 |

Any deviation of $A_1$, $A_2$, $A_3$, $A_{Ba}$ and $A_{Bc}$ from unity quantifies the level of anisotropy; the larger deviation designates higher anisotropy. The values of $A_B$ and $A_G$ are greater than zero indicating anisotropy, and a value of 100% designates its maximum anisotropy. The estimated pressure dependent values of all anisotropy indices are summarized in **Table 4**. The compound $Sn_4Au$ exhibits anisotropic nature at all pressures; however, the level of anisotropy shows nonmonotonic variation as a function of pressure. The compressibility anisotropy factors ($A_{Ba}$ and $A_{Bc}$) also designate that Sn$_4$Au is an anisotropic material.

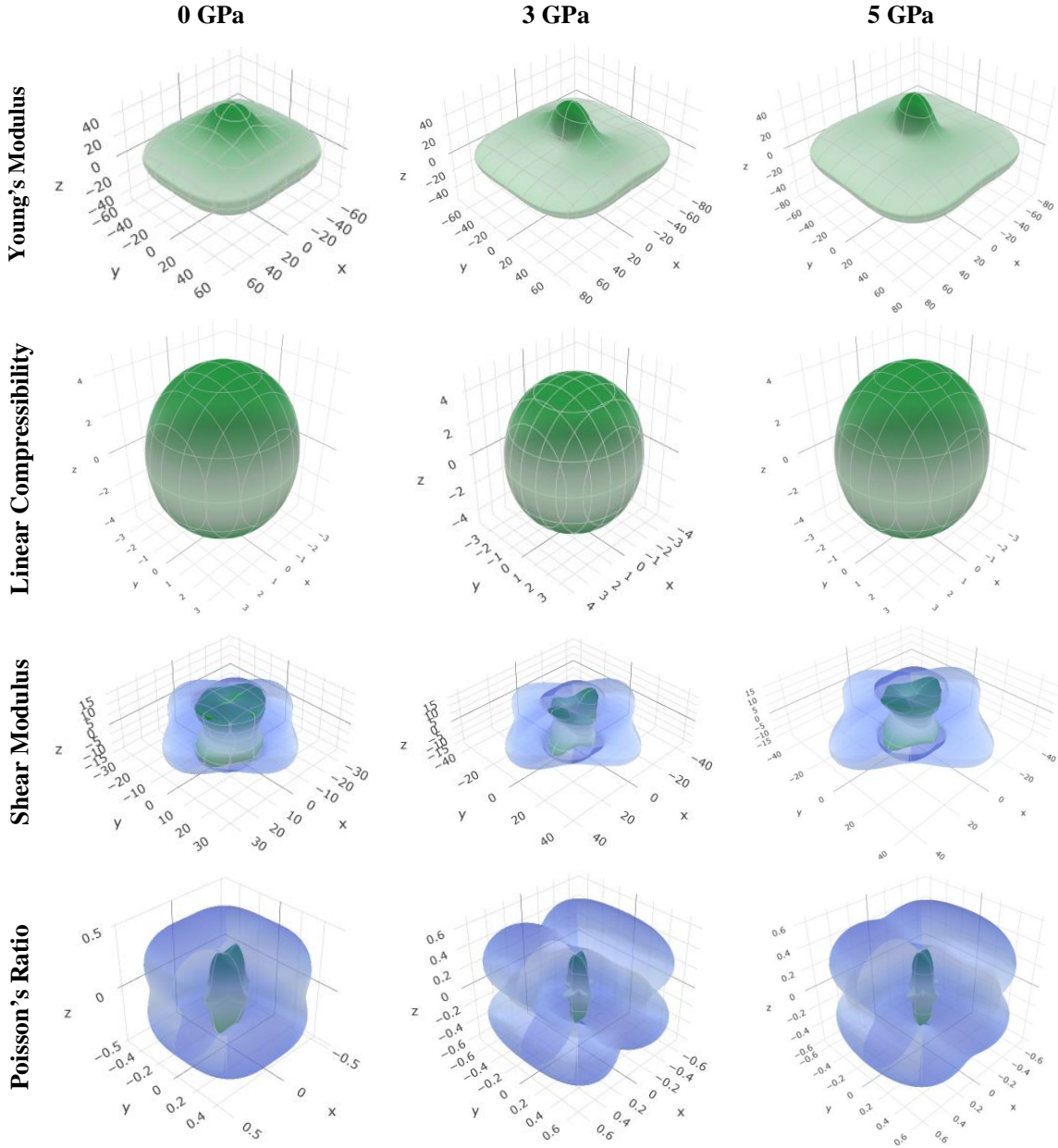

**Fig. 9**. 3D contour plots of elastic moduli and ratio [Young's modulus ($Y$), linear compressibility ($\beta$), shear modulus ($G$) and Poisson's ratio ($\sigma$)] for Sn$_4$Au at 0 GPa, 3 GPa and 5 GPa.

Three-dimensional (3D) contour plots are useful to further explore the anisotropic feature of solids in view of elastic moduli. ELATE program [93] has been used to visualize these moduli for Sn$_4$Au semimetal at representative 0 GPa, 3 GPa, and 5 GPa hydrostatic pressures. A perfect sphere represents the highest level of crystal isotropy, while any deviation from this shape indicates a lower level of isotropy [94]. **Figs. 9** depict the 3D contour plots of elastic moduli; and these plots clearly indicate their anisotropy. Moreover, the inclusion of hydrostatic pressure diminishes the level of isotropy in each case.

*3.4 Acoustic velocities and anisotropy*

Acoustic velocities, such as longitudinal ($v_l$) and transverse sound velocity ($v_t$), are essential parameters that reflect the thermal and transport properties of materials in solids. In isotropic materials, these velocities can be estimated using the shear modulus ($G$) and bulk modulus ($B$) through Navier's equation [95]:

$$v_l = \left(\frac{3B + 4G}{3\rho}\right)^{\frac{1}{2}}; \ v_t = \left(\frac{G}{\rho}\right)^{\frac{1}{2}}; \ v_m = \left[\frac{1}{3}\left(\frac{2}{v_t^3} + \frac{1}{v_l^3}\right)\right]^{-\frac{1}{3}} \quad (25)$$

where, $v_m$ and $\rho$ denote the average sound velocity and density of the crystal, respectively.

Using **Eqn. 25**, we can calculate another important physical parameter known as the Debye temperature ($\Theta_D$). This temperature is closely related with to various physical properties, such as specific heat, melting temperature, thermal conductivity, superconducting transition temperature etc. Debye temperature has been calculated using the Anderson model [96]:

$$\Theta_D = \frac{h}{k_B}\left[\frac{3n}{4\pi}\left(\frac{N_A \rho}{M}\right)\right]^{\frac{1}{3}} v_m \quad (26)$$

where, $h$ and $k_B$ represent Planck's and Boltzmann's constants, $N_A$ and $M$ denote Avogadro's number and molecular weight of $Sn_4Au$, and $n$ is the number of atoms within one molecular formula.

The estimated sound velocities ($v_l$, $v_t$, and $v_m$) alongside Debye temperatures ($\Theta_D$) of $Sn_4Au$ at 0-5 GPa pressures are summarized in **Table 5**. The pressure induced variations of sound velocities and Debye temperature are small for the pressure range considered. Moreover, the Debye temperature of the compound is also small which results from relative softness and large atomic mass of the Au atom.

**Table 5**. Estimated longitudinal, transverse and average sound velocities ($v_l$, $v_t$ and $v_m$ in km/s), Debye temperatures ($\Theta_D$ in K), acoustic Grüneisen constant ($\gamma_a$) and melting temperatures ($T_m$ in K) of Sn$_4$Au at various pressures.

| $P$ (GPa) | $v_l$ | $v_t$ | $v_m$ | $\Theta_D$ | $\gamma_a$ | $T_m$ |
|---|---|---|---|---|---|---|
| 0 | 3.109 | 1.404 | 1.583 | 159.60 | 2.328 | 1101.37 |
| 1 | 3.227 | 1.463 | 1.649 | 167.17 | 2.316 | 1160.00 |
| 2 | 3.296 | 1.456 | 1.643 | 167.41 | 2.395 | 1192.15 |
| 3 | 3.274 | 1.429 | 1.614 | 165.20 | 2.430 | 1225.38 |
| 4 | 3.396 | 1.428 | 1.615 | 166.04 | 2.540 | 1272.36 |
| 5 | 3.477 | 1.456 | 1.647 | 169.98 | 2.552 | 1307.52 |

The acoustic Grüneisen constant ($\gamma_a$) can predict the anharmonic behavior of solids and its relation to the material's thermal expansion [97]. This important parameter can be estimated as [41]:

$$\gamma_a = \frac{3}{2}\left(\frac{3v_l^2 - 4v_t^2}{v_l^2 + 2v_t^2}\right) = \frac{3(1+\sigma)}{2(2-3\sigma)} \tag{27}$$

The calculated values of $\gamma_a$ for Sn$_4$Au under different pressures (cf. **Table 5**) suggest that pressure enhances $\gamma_a$ from 2.236 to 2.552 (at 5 GPa). The $\gamma_a$ of Sn$_4$Au is quite substantial. This suggests that anharmonicity plays a significant role in the lattice dynamics of this topological semimetal.

Another crucial thermo-physical parameter is the melting temperature $T_m$. Melting temperature is used to justify whether a material is a potential candidate for high temperature applications or not [94]. $T_m$ is estimated from the formula [98] given below:

$$T_m = [553 + 5.911 C_{11}] \tag{28}$$

**Table 5** also lists the calculated values of $T_m$ for Sn$_4$Au with varying pressure. It is obvious that $T_m$ rises as the pressure increases; this is the standard behavior for solids. The melting temperature of this compound is medium, another indication of moderate bonding strength between the atoms.

The anisotropy in sound velocities can be characterized by estimating their direction-dependent values as a function of pressure within the Christoffel's equation [99]. The solution of this equation has two parts; *e.g.*, one longitudinal ($v_l$) and two transverse modes-first transverse ($v_{t_1}$) and second transverse ($v_{t_2}$). For orthorhombic Sn$_4$Au, these are computed along six principal directions [100], [010], [001], [110], [101], and [011] as follows [100]:

$$\left.\begin{aligned}
v_l^{[100]} &= \sqrt{\frac{C_{11}}{\rho}}; \quad v_{t_1}^{[100]} = v_{t_1}^{[010]} = \sqrt{\frac{C_{66}}{\rho}}; \quad v_{t_2}^{[100]} = v_{t_1}^{[001]} = \sqrt{\frac{C_{55}}{\rho}} \\
v_l^{[010]} &= \sqrt{\frac{C_{22}}{\rho}}; \quad v_{t_1}^{[010]} = v_{t_1}^{[100]} = \sqrt{\frac{C_{66}}{\rho}}; \quad v_{t_2}^{[010]} = v_{t_2}^{[001]} = \sqrt{\frac{C_{44}}{\rho}} \\
v_l^{[001]} &= \sqrt{\frac{C_{33}}{\rho}}; \quad v_{t_1}^{[001]} = v_{t_2}^{[100]} = \sqrt{\frac{C_{55}}{\rho}}; \quad v_{t_2}^{[001]} = v_{t_2}^{[010]} = \sqrt{\frac{C_{44}}{\rho}} \\
v_l^{[110]} &= \sqrt{\left(\frac{C_{44} + C_{55}}{2\rho}\right)}; \quad v_{t_1}^{[110]} = v_{t_2}^{[110]} = \sqrt{\left(\frac{C_{11} + C_{22} + 2C_{66}}{4\rho}\right)} \\
v_l^{[101]} &= \sqrt{\left(\frac{C_{44} + C_{66}}{2\rho}\right)}; \quad v_{t_1}^{[101]} = v_{t_2}^{[101]} = \sqrt{\left(\frac{C_{11} + C_{33} + 2C_{55}}{4\rho}\right)} \\
v_l^{[011]} &= \sqrt{\left(\frac{C_{55} + C_{66}}{2\rho}\right)}; \quad v_{t_1}^{[011]} = v_{t_2}^{[011]} = \sqrt{\left(\frac{C_{22} + C_{33} + 2C_{44}}{4\rho}\right)}
\end{aligned}\right\} \quad (29)$$

**Table 6.** The calculated pressure-dependent anisotropic sound velocities for orthorhombic $Sn_4Au$ semimetal.

| $P$ (GPa) | [100] | | | [010] | | | [001] | | | [110] | | [101] | | [011] | |
|---|---|---|---|---|---|---|---|---|---|---|---|---|---|---|---|
| | $v_l$ | $v_{t_1}$ | $v_{t_2}$ | $v_l$ | $v_{t_1}$ | $v_{t_2}$ | $v_l$ | $v_{t_1}$ | $v_{t_2}$ | $v_l$ | $v_{t_1} = v_{t_2}$ | $v_l$ | $v_{t_1} = v_{t_2}$ | $v_l$ | $v_{t_1} = v_{t_2}$ |
| 0 | 3.278 | 1.970 | 0.975 | 3.277 | 1.970 | 1.192 | 3.041 | 0.975 | 1.192 | 1.089 | 2.419 | 1.628 | 2.340 | 1.554 | 2.389 |
| 1 | 3.421 | 2.061 | 1.070 | 3.386 | 2.061 | 1.173 | 3.147 | 1.070 | 1.173 | 1.123 | 2.526 | 1.677 | 2.444 | 1.642 | 2.456 |
| 2 | 3.485 | 2.090 | 0.977 | 3.500 | 2.090 | 1.202 | 3.232 | 0.977 | 1.202 | 1.095 | 2.578 | 1.705 | 2.475 | 1.631 | 2.529 |
| 3 | 3.549 | 2.135 | 1.030 | 3.544 | 2.135 | 1.193 | 2.953 | 1.030 | 1.193 | 1.115 | 2.618 | 1.729 | 2.421 | 1.676 | 2.456 |
| 4 | 3.648 | 2.187 | 0.823 | 3.623 | 2.187 | 1.185 | 3.357 | 0.823 | 1.185 | 1.020 | 2.698 | 1.759 | 2.546 | 1.652 | 2.608 |
| 5 | 3.714 | 2.236 | 0.905 | 3.708 | 2.236 | 1.108 | 3.432 | 0.905 | 1.108 | 1.012 | 2.758 | 1.765 | 2.608 | 1.706 | 2.645 |

The computed sound velocities under pressure along different directions are listed in **Table 6**. It is seen that there are two pure transverse modes along [100], [010], and [001] directions, and two degenerate transverse wave modes along the [110], [101], and [011] directions. The highest velocities at 0 GPa are found to be 3.278 km/s and 3.277 km/s in longitudinal mode along [001] and [010] directions, respectively, owing to the maximum values of $C_{11}$ and $C_{22}$ among all $C_{ij}$'s. This finding is very much allied with the literature [101]. Therefore, longitudinal sound velocity follows the sequence [100] > [010] > [001] > [101] > [011] > [110], and consequently, $Sn_4Au$ manifest anisotropic features in sound propagation owing to their varying directional atomic arrangements and bondings.

### 3.5 Thermal conductivities and anisotropy

#### 3.5.1 Minimum thermal conductivities and anisotropy

The limiting value of thermal conductivity, known as minimum thermal conductivity ($\kappa_m$), is an intrinsic property of matter which has a practical importance for high temperature applications [102]. To evaluate $\kappa_m$, we have used two separate models: Cahill's model [103] and Clarke's model [104] in which Cahill's model includes both the longitudinal and transverse acoustic modes, as follows [103]:

$$k_m^{Cahill} = \begin{cases} \dfrac{1}{2.48} k_B n_v^{\frac{2}{3}} (v_l + 2v_t) & \text{(Isotrpoic)} \\ \dfrac{1}{2.48} k_B n_v^{\frac{2}{3}} (v_l + v_{t_1} + v_{t_2}) & \text{(Anisotrpoic)} \end{cases} \quad (30)$$

$$k_m^{Clarke} = k_B v_m \left(\dfrac{M}{n\rho N_A}\right)^{-\frac{2}{3}} \quad (31)$$

where $k_B$ and $n_v$ represent Boltzmann's constant and the number of atoms per unit volume, respectively. The thermal conductivities, whether isotropic or anisotropic, are estimated using **Eqn. 30**. In the equation, the two shear wave velocities are assumed to be identical for isotropic solids.

**Table 7**. Computed isotropic ($\kappa_m$ in Wm$^{-1}$K$^{-1}$ from both Cahill and Clarke method) and anisotropic minimum thermal conductivities ($\kappa_m$ in Wm$^{-1}$K$^{-1}$ from Cahill method) and lattice thermal conductivities ($\kappa_l$ at 300 K in Wm$^{-1}$K$^{-1}$) for Sn$_4$Au at various pressures.

| $P$ (GPa) | $k_m^{Clarke}$ | $k_m^{Cahill}$ | $k_m^{Cahill}$ along | | | | | | $\kappa_l$ |
|---|---|---|---|---|---|---|---|---|---|
| | | | [100] | [010] | [001] | [110] | [101] | [011] | |
| 0 | 0.630 | 0.377 | 0.396 | 0.410 | 0.332 | 0.377 | 0.402 | 0.403 | 3.163 |
| 1 | 0.663 | 0.396 | 0.422 | 0.426 | 0.347 | 0.397 | 0.423 | 0.422 | 3.657 |
| 2 | 0.667 | 0.403 | 0.426 | 0.442 | 0.352 | 0.406 | 0.433 | 0.435 | 3.399 |
| 3 | 0.662 | 0.402 | 0.441 | 0.451 | 0.340 | 0.417 | 0.431 | 0.432 | 3.149 |
| 4 | 0.668 | 0.414 | 0.441 | 0.463 | 0.355 | 0.424 | 0.453 | 0.454 | 2.893 |
| 5 | 0.686 | 0.426 | 0.457 | 0.470 | 0.363 | 0.435 | 0.466 | 0.467 | 3.061 |

**Table 7** summarizes the isotropic and anisotropic minimum thermal conductivities along principal directions at various pressures. The computed thermal conductivities clearly indicate anisotropic features due to their dissimilar sound velocities along different directions. Moreover, inclusion of pressure causes an increase in the minimum thermal conductivity in a non-monotonic fashion. The trend in variation of $\kappa_m$ with pressure is in agreement with that of Debye temperature. This trend validates Callaway–Debye theory [105]: lower thermal conductivity is associated with a smaller Debye temperature. The computed thermal conductivity of Sn$_4$Au is low and the compound can be useful as thermal insulator.

*3.5.2 Lattice thermal conductivity*

Lattice thermal conductivity ($\kappa_l$) generates an idea about the anharmonicity of a structure and can predict the potentiality of materials to be used in thermoelectric applications. The large value of $\kappa_l$ quantifies the presence of significant anharmonic effects, and therefore diminishes the figure-of-merit (*ZT*) of solid as $\kappa_l$ is inversely proportional to *ZT* [106]. Temperature-dependent $\kappa_l$ is evaluated from the expression by Slack [107] as follows:

$$k_l = A \frac{M_{av} \Theta_D^3 \delta}{\gamma_a^2 n^{\frac{2}{3}} T} \tag{32}$$

where, $M_{av}$ (in kg/mol) denotes the average atomic mass, $\delta$ represents the cubic root of the average atomic volume in meter, and $\gamma_a$ is the acoustic Grüneisen constant, and the factor $A\ (\gamma_a)$ can be calculated according to Julian [108] as:

$$A(\gamma_a) = \frac{2.4281 \times 10^7}{(1 - 0.514\gamma_a^{-1} + 0.228\gamma_a^{-2})} \tag{33}$$

The calculated lattice thermal conductivity for pressures up to 5 GPa at 300 K is shown in **Table 7**. The variation of lattice thermal conductivity with pressure is non-monotonic with the highest value obtained under a pressure of 1 GPa.

### *3.6 Electronic properties*

#### *3.6.1 Band structure and density of states*

To get deep insights into the electronic nature, the bulk electronic band structure, total density of states (TDOS) and partial density of states (PDOS) for $Sn_4Au$ are calculated in this study. The band structure calculations for $Sn_4Au$ is carried out without considering spin-orbit coupling (SOC) effect as Karn *et al*. [36] found insignificant effect when considering SOC. The electronic energy dispersion curves of $Sn_4Au$ along high symmetry points ($\Gamma$-Z-T-Y-S-X-U-R) of the BZ in the energy range from −2 eV to +2 eV at 0 GPa, 3 GPa and 5 GPa pressures are depicted in **Figs. 10a-c**. The Fermi level ($E_F$) is set to zero energy throughout the calculations. Important electronic bands (band indices are given in the figure), including those crossing the Fermi level are shown in different colors for 0 GPa, 3 GPa and 5 GPa. There is no bandgap in the energy dispersion and therefore, the compound is metallic for the pressures considered. The band structure at 0, 3, and 5 GPa are quite similar. The behavior of electronic bands in the vicinity of the Fermi level makes topological semimetal (TS) different from the ordinary metals. The inset of **Figs. 10** clearly indicates topological signature just below the Fermi energy when we generate a projected view of energy axis from -0.7 eV to 0.3 eV in **Fig. 10a** and **Fig. 10c** while from -0.5 eV to 0.1 eV for **Fig. 10b**. Moreover, Dirac cone type semi-metallic features appeared in the band structure for each pressure (cf. **Figs. 10**). A detailed features of Dirac type semimetal along with fundamental topological behaviors have been studied previously by Karn *et al*. [36]. From the curvatures of the bands crossing the Fermi level, both electron- and hole-like Fermi surfaces are expected within the BZ. The largest contribution to the Fermi surface comes from band number 56 (purple) as seen in **Figs. 10**.

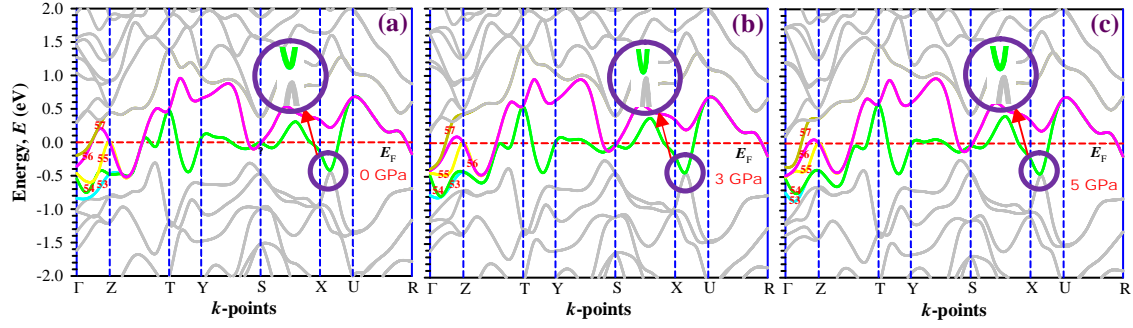

**Fig. 10**. The electronic band structures of $Sn_4Au$ for (a) 0 GPa, (b) 3 GPa and (c) 5 GPa pressures. Inset shows the signature of topological dispersion.

To further elucidate the electronic nature of $Sn_4Au$, its TDOS and PDOS are calculated and depicted in **Fig. 11**. The upper panel represents the TDOS curve that clearly indicates the metallic/semi-metallic nature as the density of states is nonzero at $E_F$. In the lower energy regime from about -10.0 eV to -6.0 eV, Sn-$s$ electrons are the foremost contributors to the TDOS alongside a little contribution from Sn-$p$ electrons. A strong hybridization of Au-$d$ and Sn-$p$ orbitals appear in the energy range from -6.2 eV to -3.8 eV and results in a large peak in the TDOS. However, the main contribution of bands near $E_F$ corresponds to $s$, $p$-orbitals of Sn and $p$, $d$-orbitals of Au atoms.

The projected view of TDOS at various pressures within the energy range -1.0 eV to +1.0 eV is represented in **Fig. 12a** and the variation of DOS at $E_F$ with pressure is depicted in **Fig. 12b**. It is evident from these figures that hydrostatic pressure diminishes the DOS at $E_F$. This is because the bands close to Fermi energy becomes more dispersive as pressure increases.

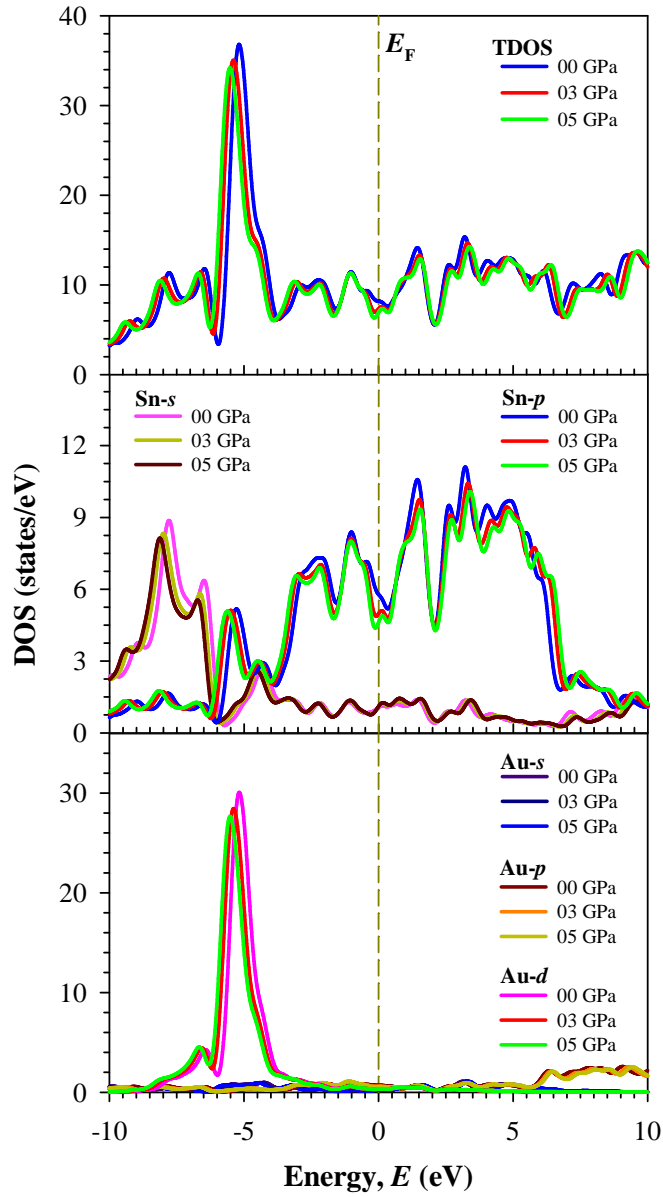

**Fig. 11**. Calculated TDOS and PDOS of Sn$_4$Au at various pressures.

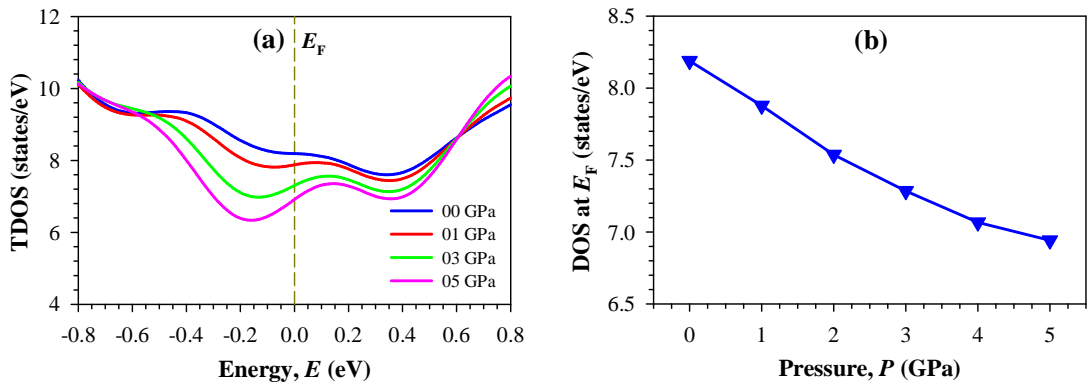

**Fig. 12**. (a) TDOS near $E_F$ and (b) DOS at $E_F$ as a function of pressure for Sn$_4$Au.

*3.6.2 Mulliken population and bonding characters*

Mulliken's atomic population (MAP) analysis is useful to explain bond overlap population (BOP) for neighboring atoms and the effective valence charge (EVC). Zero/negative BOP signifies either negligible interactions between atoms or emphasize anti-bonding levels, and therefore exhibits a remarkably weak bonding that might be ineffective in estimating theoretical hardness [109]. A large/small positive value of BOP, conversely, indicates the high degree of covalency/iconicity. EVC, the difference between the formal ionic charge and Mulliken/Hirshfeld charge within a material, predicts the bonding nature of compounds. Although the overall features of the bonding nature often remain unaltered, the magnitude of Mulliken charge is generally higher than that of Hirshfeld charge. Consequently, the strength of both covalency and ionicity is lower in the Hirshfeld population analysis (HPA) [110]. The ideal ionic bonding appears when the value of EVC is zero, and any departure from zero signifies the strength of covalent bonding [111].

**Table 8**. Calculated Mulliken atomic populations, and effective valence charge (EVC), Mulliken bond number $n^\mu$, bond overlap population $P^\mu$, bond length $d^\mu$, bond volume $v_b^\mu$, bond hardness of $\mu$-type $H_v^\mu$, and hardness $H$ of $Sn_4Au$ at various pressures (P in GPa).

| P | Mulliken atomic populations | | | | | | | | | Mulliken bond overlap population | | | | | Hardness |
|---|---|---|---|---|---|---|---|---|---|---|---|---|---|---|---|
| | Species | s | p | d | Total | Mulliken Charge | Formal Charge | EVC | Hirshfeld Charge | EVC | Bond | $n^\mu$ | $P^\mu$ | $d^\mu$ | $v_b^\mu$ | $H_v^\mu$ | H |
| 0 | Sn | 1.58 | 2.16 | 0.00 | 3.74 | 0.26 | +4 | 3.74 | -0.01 | 3.99 | Sn–Au (I) | 8 | 0.14 | 2.8842 | 15.821 | 1.03905 | 0.874 |
| | | | | | | | | | | | Sn–Au (II) | 8 | 0.13 | 2.8917 | 15.943 | 0.95256 | |
| | Au | 1.16 | 1.28 | 9.60 | 12.04 | -1.04 | +3 | 1.96 | 0.03 | 2.97 | Sn–Au (III) | 8 | 0.11 | 2.9191 | 16.402 | 0.76877 | |
| | | | | | | | | | | | Sn–Au (IV) | 8 | 0.11 | 2.9202 | 16.421 | 0.76729 | |
| 3 | Sn | 1.55 | 2.19 | 0.00 | 3.74 | 0.26 | +4 | 3.74 | -0.01 | 3.99 | Sn–Au (I) | 8 | 0.14 | 2.8497 | 15.080 | 1.1255 | 0.915 |
| | | | | | | | | | | | Sn–Au (II) | 8 | 0.14 | 2.8499 | 15.084 | 1.1251 | |
| | Au | 1.14 | 1.31 | 9.59 | 12.05 | -1.05 | +3 | 1.95 | 0.04 | 2.96 | Sn–Au (III) | 8 | 0.10 | 2.8936 | 15.788 | 0.7447 | |
| | | | | | | | | | | | Sn–Au (IV) | 8 | 0.10 | 2.8943 | 15.800 | 0.7438 | |
| 5 | Sn | 1.53 | 2.20 | 0.00 | 3.74 | 0.26 | +4 | 3.74 | -0.01 | 3.99 | Sn–Au (I) | 8 | 0.14 | 2.8299 | 9.350 | 2.4967 | 2.540 |
| | | | | | | | | | | | Sn–Au (II) | 8 | 0.14 | 2.8306 | 9.357 | 2.4936 | |
| | | | | | | | | | | | Sn–Au (III) | 8 | 0.09 | 2.8755 | 9.809 | 1.4818 | |
| | Au | 1.14 | 1.32 | 9.59 | 12.05 | -1.05 | +3 | 1.95 | 0.04 | 2.96 | Sn–Au (IV) | 8 | 0.09 | 2.8803 | 9.858 | 1.4695 | |
| | | | | | | | | | | | Sn–Au (V) | 8 | 0.35 | 2.9700 | 10.808 | 4.9023 | |
| | | | | | | | | | | | Sn–Au (VI) | 8 | 0.30 | 2.99378 | 11.0699 | 4.0375 | |

The calculated results regarding MAP and EVC of $Sn_4Au$ at different pressures are disclosed in **Table 8**. It is noteworthy that 'Au' atom bears negative charges and 'Sn' bears positive charges in $Sn_4Au$. Thus the charge is transferred from 'Sn' cation towards 'Au' anion, suggesting an ionic contribution to the Sn-Au bonding. The amount of atomic charges of 'Sn' and 'Au' atom are 0.26*e* and -1.04*e*, respectively for 0 GPa. Both deviates from the formal value from a purely ionic state (Sn: +4 and Au: +3); this deviation also reflects the covalent bonding feature between the atomic species. The calculated results of EVC of Sn and Au in $Sn_4Au$ are +3.74 and +1.96, respectively also stipulate the investigated compound as partly covalent. The small positive values of BOP within different bonds of Sn-Au signify the weak bondings between the atoms, confirming its soft nature. Therefore, the bonding in this compound is hybrid; both ionic and covalent nature contributes.

*3.6.3 Bond hardness*

Theoretical bond hardness $H_v$, one of the prominent mechanical parameter, can be estimated within the Gao model [112] which is applicable for complex multiband solids based on bond length and BOP, as [113]:

$$H_v = \left[\prod^{\mu}(H_v^{\mu})^{n^{\mu}}\right]^{1/\Sigma n^{\mu}} \qquad (34)$$

$$H_v^{\mu} = 740\, P^{\mu}(v_b^{\mu})^{-5/3} \qquad (35)$$

$$v_b^{\mu} = (d^{\mu})^3 \Big/ \sum_v [(d^v)^3 N_b^v] \qquad (36)$$

where $H_v^{\mu}$ and $n^{\mu}$ represent the hardness and bond number of $\mu$-type bonds, respectively, $P^{\mu}$ and $v_b^{\mu}$ denote bond overlap population and bond volume of $\mu$-type, respectively, while $d^{\mu}$ and $N_b^v$ indicates the bond length and bond number of $v$-type per unit volume, respectively. The calculated bond hardness and average hardness at various pressures are listed in **Table 8**. The hardness are found to be 0.874 GPa, 0.915 GPa and 2.540 GPa for 0 GPa, 3 GPa and 5 GPa, respectively, indicating that the maximum hardness is obtained at 5 GPa in support with their elastic moduli and charge density distribution (discussed in the next section).

*3.6.4 Charge density and bonding characters*

The electronic charge density (CD) distribution is a favorable tool to acquire an insightful understanding of chemical bonding between the ions. A CD mapping is more effective than electrostatic potential (ESP) mapping as it is less sensitive to long-range electrostatic effects. The calculated charge density distribution (CDD) maps of $Sn_4Au$ in (010) and (110) planes under different hydrostatic pressures are shown in **Figs. 13**. The color scales on the right-hand side of each EDD map illustrate the total electron density in e/Å$^3$. The accumulation of charges between two atoms or non-spherical CDD around atoms indicates the covalent bonding, whereas the balancing of positive or negative charge at the atomic positions indicates ionic bonding. Moreover, uniform charge smearing illustrates metallic bonding [84]. The CDD around Sn atoms is non-spherical whereas Au atoms exhibit spherical charge distribution. The CDD map shows weak signatures of covalent bonding between Sn-Sn and Sn-Au atoms at 0 GPa. The maximum charges are accumulated around the core region of Au atoms for both planes. The charge accumulation between atoms increases with applied pressure. Therefore, the covalent bonding between atoms becomes stronger with pressure. The directional and plane dependency of charge density distribution is clearly observable. With applied pressures of 3 GPa and 5 GPa, the CDD between the atoms increases and the directional anisotropy is maintained throughout. Both the Mulliken bond population (cf.

section 3.6.2) and electronic DOS (cf. section 3.6.1) studies are in consistent alignment with these findings.

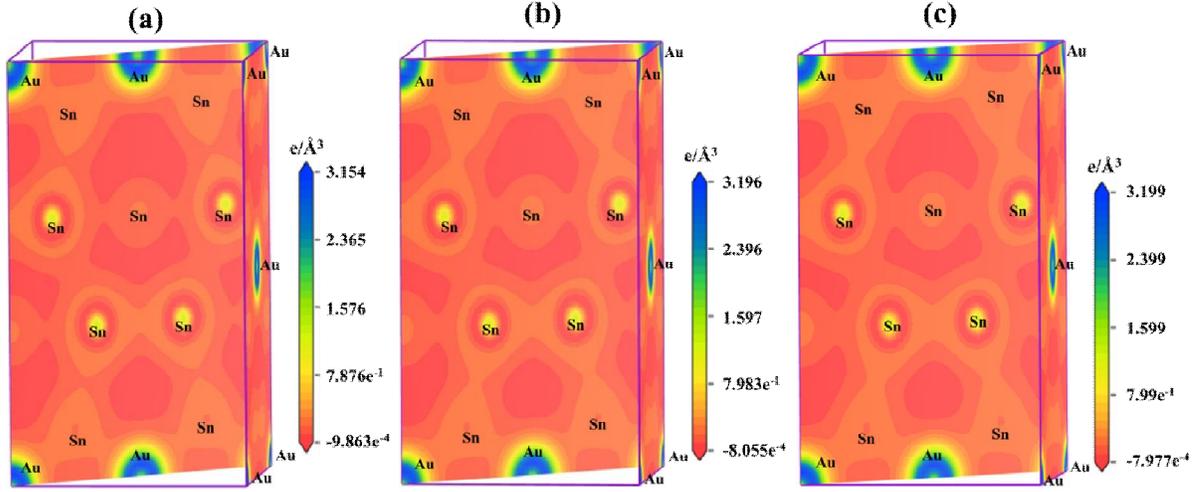

**Fig. 13.** CDD maps of $Sn_4Au$ in the (010) and (110) planes under (a) 0 GPa, (b) 3 GPa, and (c) 5 GPa.

*3.7 Optical properties*

The optical properties of a material play a pivotal role to predict whether it can be used for optoelectronic and photovoltaic device applications or not. In this study, the various optical functions, namely, dielectric function, refractive index, reflectance, optical conductivity, absorption coefficient and electron energy loss function of $Sn_4Au$ have been calculated for photon energies up to 20 eV under pressures within 0 GPa - 5 GPa. As the material under study is anisotropic, all calculations are done for two polarization directions, <100> and <001>, of the electric field vector. Due to the metallic nature of $Sn_4Au$, a semi-empirical Drude term with Gaussian smearing are utilized to compute the optical parameters.

The complex dielectric function, $\varepsilon(\omega)$ is the key to describe the pathway in which a material responds to electromagnetic radiation in the infrared (IR)-visible to ultraviolet (UV) regions. Generally, the real component $\varepsilon_1(\omega)$ elucidates the dispersion and polarization of light in materials, meanwhile the imaginary part $\varepsilon_2(\omega)$ reveals the absorption of light. At lower energies (IR region), the optical spectra is due to the intraband transitions of electrons in metallic compounds whereas the interband transitions are strongly dependent on the electronic band structure [114-117]. The calculated pressure-dependent real and imaginary parts of the dielectric function for $Sn_4Au$ is illustrated in **Fig. 14** at three different representative pressures (0, 3, and 5 GPa).

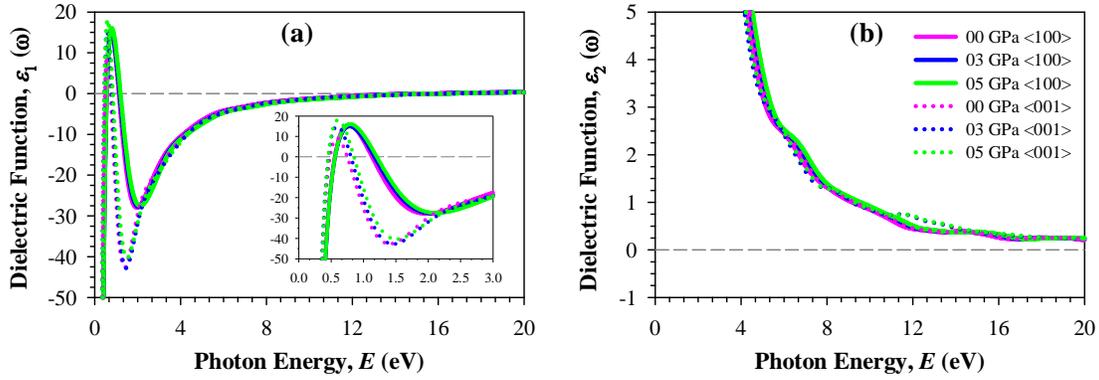

**Fig. 14.** The (a) real part $\varepsilon_1$ and (b) imaginary part $\varepsilon_2$ of the dielectric function of $Sn_4Au$ under different pressures.

Characteristic peaks of $\varepsilon_1(\omega)$ for <100> polarization are attained at ~ 0.77 eV, 0.79 eV, and 0.80 eV for 0 GPa, 3 GPa and 5 GPa, respectively, in the IR region (**Fig. 14a**). In <001> direction, on the other hand, maxima of $\varepsilon_1(\omega)$ occurs at lower energies at 0.59 eV, 0.60 eV and 0.61 eV for 0 GPa, 3 GPa and 5 GPa, respectively. The peak heights also show small polarization direction dependence. After reaching the peak, a sharp decrease with rising energy appears and $\varepsilon_1(\omega)$ becomes negative from 1.11 eV to 15.7 eV for <100> and from 0.76 eV to 16.03 eV at 0 GPa for <001> polarizations, respectively (Inset of **Fig. 14a** clarifies the scenario). The imaginary part of the dielectric function is shown in **Fig. 14b**. It is observed from **Fig. 14b** that no peak in $\varepsilon_2(\omega)$ appeared in the lower energy region up to ~ 4 eV and $\varepsilon_2(\omega)$ falls featurelessly up to 20 eV. Optical anisotropy in $\varepsilon_2(\omega)$ is also very weak.

The complex refractive index $N(\omega)$ [$= n(\omega) + ik(\omega)$] comprises two parts: refractive index $n(\omega)$ analyzes the phase velocity of the photons in the medium and the extinction coefficient $k(\omega)$ elucidates the energy loss of the electromagnetic wave in the medium which is directly correlated to the dielectric constant and absorption coefficient [116,117]. $n(\omega)$ and $k(\omega)$ as a function of photon energy within 0 to 20 eV is illustrated in **Fig. 15a-b** (inset shows their significant peaks). The refractive index exhibits the prime peaks in the IR region which gradually diminishes in the visible-to-UV region, as seen in **Fig. 15a**. In the IR and visible region, the refractive index is anisotropic. At higher energies, this anisotropy vanishes. The real part $n(\omega)$ is quite high in the low energy region. High refractive index materials can be integrated into infrared detectors and used in designing optical devices to achieve improved light distribution and increased brightness [118]. The extinction coefficient $k(\omega)$, on the other hand, after reaching its maxima in the IR region, decreases in the visible to UV region as seen in **Fig. 15b**. The effect of pressure is weak.

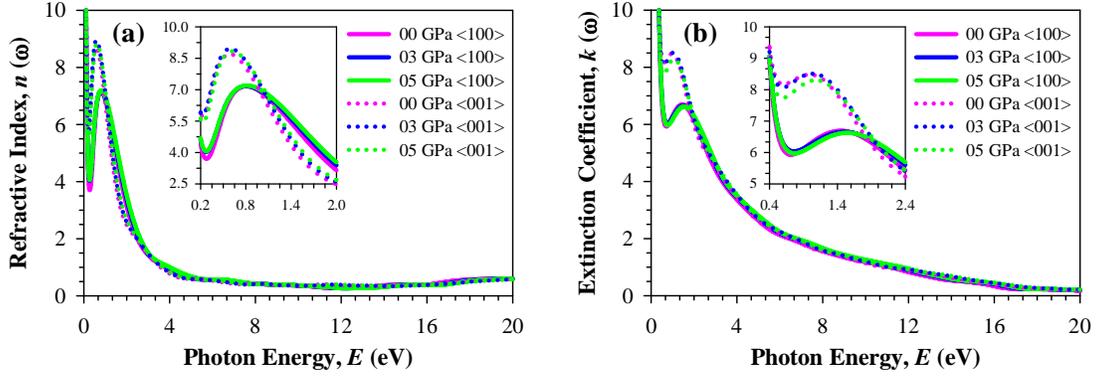

**Fig. 15**. Calculated (a) refractive index $n$ and (b) extinction coefficient $k$ of $Sn_4Au$ under different pressures as a function of photon energy.

Energy-dependent absorption coefficient $α(ω)$ of $Sn_4Au$ is depicted in **Fig. 16a**. As seen, the absorption spectra begins at zero photon energy as a signature of metallic compounds [119]. Generally, high absorbance solids are widely used in optoelectronic devices in both visible and UV regions [119]. Optical anisotropy is low and the effect of pressure on the absorption coefficient is weak as well.

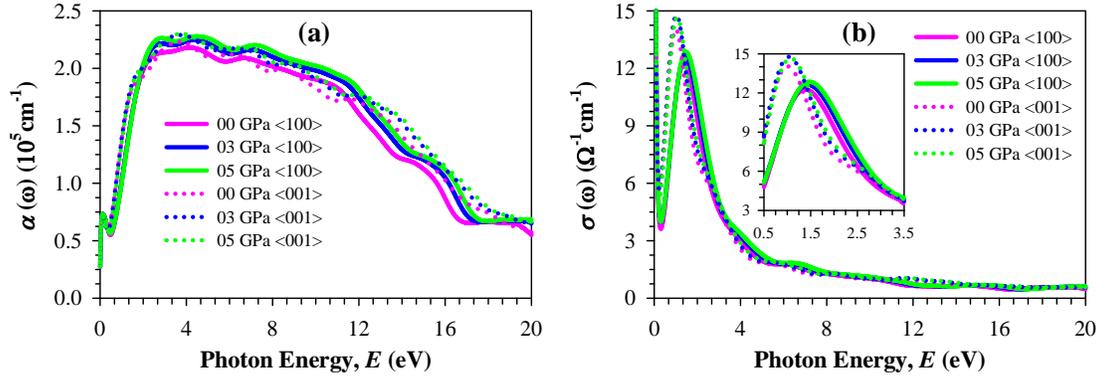

**Fig. 16**. Pressure-dependent (a) absorption coefficient $α$ and (b) optical conductivity $σ$ as a function of photon energy for $Sn_4Au$.

The real part of optical conductivity, plotted in **Fig. 16b** (inset shows significant peaks), exhibits sharp dip in the energy range 0~0.3 eV, and thereafter rises to maxima in the energy range 1.3–1.5 eV and 0.9–1.1 eV for <100> and <001> polarizations, respectively. Optical conductivity spectra demonstrates the change in dynamical conductivity as a function of photon energy. The photoconductivity begins from zero photon energy which signifies the conductive nature of $Sn_4Au$. There is some optical anisotropy in the visible region. The anisotropy disappears in the ultraviolet region.

The reflectivity $R(ω)$ starts with a value of ~ 99% at zero photon energy and decreases to ~ 72% (IR region), and rises to maximum values of 78.0% at 2.60 eV, 78.4% at 2.77 eV and 78.6% at 2.90 eV for 0, 3 and 5 GPa, respectively for <100> polarization as seen in **Fig. 17a**. The spectra is almost identical for the <001> polarization. The reflectivity is nonselective in the visible region and stays above 75% throughout. This suggests that the compound $Sn_4Au$ can be used as an efficient reflector of visible light and can be employed as a coating material to reduce solar heating [120]. Furthermore, due to nearly constant reflectivity in the visible regime, the investigated compound might be appear as metallic white [119]. Application of pressure has minimal effect on the $R(\omega)$.

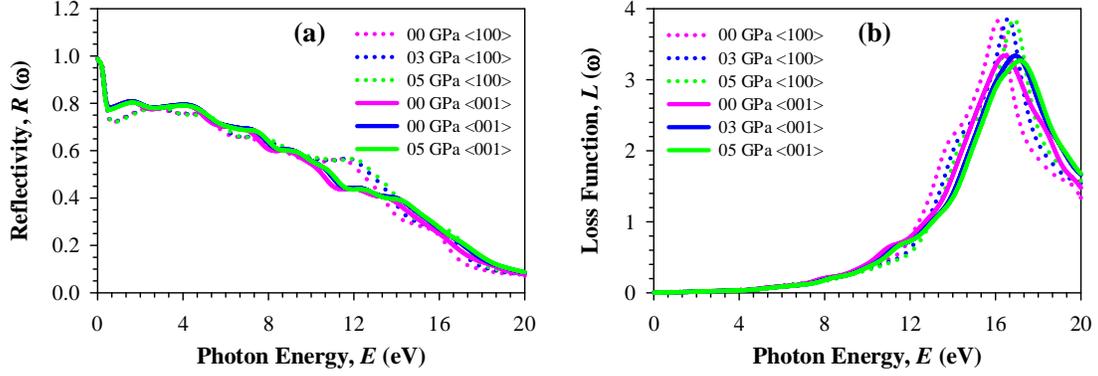

**Fig. 17**. Variation of (a) reflectivity, $R$ and (b) energy loss function, $L$ with photon energy of $Sn_4Au$ under different pressures.

The loss function $L(\omega)$ estimates the probable energy losses experienced by electrons as they interact with the material due to excitation of the plasma oscillation [121]. The calculated pressure-dependent $L(\omega)$ of $Sn_4Au$ are shown in **Fig. 17b**. This function may easily be described in view of the reciprocal of the imaginary part of complex dielectric function. That is, the loss functions and the imaginary dielectric functions are negatively correlated. The dielectric part gets close to zero when the peak of $L(\omega)$ appears [121]. The energy loss spectra correspond to the frequency of collective oscillations of the valence electrons. The associated peaks are closely related to the plasma oscillation and the characteristic frequency/energy is referred to as bulk plasma frequency ($\omega_p$) [122,123]. From **Fig. 17b**, we see that the peaks in $L(\omega)$ occur at 16.11 eV, 16.60 eV, and 16.86 eV for 0, 3, and 5 GPa, respectively, when the electric field polarization is along <100> direction. For the <001> direction, the peaks are at 16.43 eV, 16.95 eV, and 17.15 eV for 0, 3, and 5 GPa, respectively. Both reflectivity and absorption coefficient fall sharply at the plasma frequency. This indicates that the compound under investigation becomes transparent to electromagnetic wave above the plasma energy [124]. There is notable optical anisotropy in the peak feature of the loss function.

*3.8 Superconducting state properties of $Sn_4Au$ under pressure*

It has been reported that $Sn_4Au$ exhibits low-$T_c$ superconductivity with superconducting transition temperature in the range of 2.4 K to 2.6 K [25,35]. This is an example of superconductivity in a topological semimetal [25,35]. The superconducting $T_c$ of such systems can be obtained from the widely used formula proposed by McMillan [125]:

$$T_c = \frac{\theta_D}{1.45} exp\left[-\frac{1.04(1 + \lambda_{ep})}{\lambda_{ep} - \mu^*(1 + 0.62\lambda_{ep})}\right] \qquad (37)$$

From this expression for critical temperature, it becomes clear that superconducting $T_c$ depends on three physical parameters, namely, the Debye temperature, electron-phonon coupling constant ($\lambda_{ep}$), and the repulsive Coulomb pseudopotential ($\mu^*$). For the pressure range considered, the pressure dependent variation in the Debye temperature of $Sn_4Au$ is quite weak (**Table 5**). The electronic energy density of states at the Fermi level [$N(E_F)$], decreases gradually with rising pressure (**Fig. 12a**). The repulsive Coulomb pseudopotential which diminishes $T_c$, on the other hand, can be estimated from the density of states at the Fermi level [81,126]. The computed values of $\mu^*$ are: 0.174, 0.172, 0.169, 0.168, 0.166 and 0.164 for 0, 1, 2, 3, 4, and 5 GPa, respectively. This parameter is also a measure of electronic correlations. The computed values suggests that the strength of electronic correlation decreses slowly with the application of pressure in $Sn_4Au$.

For full analytic calcualtion of the pressure dependence of $T_c$, information regarding $\lambda_{ep}$ is required. Unfortunately, such calculation fall outside the scope of the CASTEP code. At the same time, for a given average electron-phonon interaction energy, $V_{ep}$, $\lambda_{ep}$ varies linearly with the value of $N(E_F)$ due to the relation, $\lambda_{ep} = N(E_F)V_{ep}$ [42,127]. This implies that $\lambda_{ep}$ might decrease with increasing pressure for $Sn_4Au$ (**Fig. 12a**). Overall, the small increase in the Debye temperature and moderate decrease in $\mu^*$ should favor an enhancement of $T_c$ with rising pressure. But the decrease in $N(E_F)$ works in an opposite way. Therefore, our work predicts a weak pressure dependent change in $T_c$ for $Sn_4Au$ within the pressure range considered.

## 4. Conclusion

Pressure-dependent physical properties of the topological semimetal $Sn_4Au$ have been investigated in this work via first-principles study within the DFT. The calculated structural parameters in the ground state are in fair agreement with the prior results. The negative formation energy (*i.e.*, $E_f < 0$), mechanical and dynamical stability, indicate that $Sn_4Au$ is thermodynamically stable. $Sn_4Au$ is ductile and highly machinable in the pressure range considered. The compound is elastically anisotropic and relatively soft in nature. The bonding character is mixed with ionic, covalent, and metallic contributions. There is significant anharmonic contribution in the phonon dynamics of $Sn_4Au$; both anharmonicity and the melting temperature increases with increasing pressure. The lattice thermal conductivity of the compound is low at all pressures considered. The electronic band structure exhibit the semimetallic nature with topological signatures. The main contributors to the TDOS at the Fermi level are the *s*, *p*-orbitals of Sn atom. The optical parameters show small anisotropy at all the pressures. The optical spectra show metallic feature and correspond well to the TDOS profile. The characteristic peaks in refractive index, reflectivity, photoconductivity, and loss function exhibit a slight shift towards the higher energy with rising pressure. The compound $Sn_4Au$ can be integrated within the infrared detectors due its fairly high refractive index in the IR region. The compound can be used to reduce solar heating because the reflectivity remains above 75% in the visible region. The compound is also an efficient absorver of UV light. We have investigated qualitatively the pressure dependence of the superconducting transition temperature of Sn$_4$Au. In the pressure range adopted, a weak variation in $T_c$ is forecasted.

To summarize, most of the results presented in this work are novel. The topological semimetal Sn$_4$Au possesses several features suitable for applications. We hope that the results presented herein will encourage researchers to explore this compound further both theoretically and experimentally.

**Acknowledgements**

S.H.N. acknowledges the research grant (1151/5/52/RU/Science-07/19-20) from the Faculty of Science, University of Rajshahi, Bangladesh, which partly supported this work. M.A.H.S. acknowledges the fellowship from the Bangabandhu Science and Technology Fellowship Trust, Ministry of Science and Technology, Bangladesh for his Ph.D. research. We dedicate this work in memory of the martyrs of the July-August 2024 revolution in Bangladesh.

**Data availability**

Data will be made available from the corresponding author on reasonable request.

**Declaration of competing interest**

The authors declare that they have no known competing financial interests or personal relationships that could have appeared to influence the work reported in this paper.

**CRediT authorship contribution statement**

**M.A.H. Shah**: Investigation, Software, Methodology, Data curation, Visualization, Formal analysis, Writing–original draft. **M.I. Naher:** Methodology, Formal analysis. **S.H. Naqib**: Conceptualization, Supervision, Project administration, Validation, Formal analysis, Resources, Writing–review & editing.